\documentclass[runningheads]{llncs}
\usepackage{graphicx} 
\usepackage{hyperref}
\usepackage{tikz}
\usepackage{fancyhdr} 
\usepackage{subcaption}
\usepackage{algpseudocode}
\usepackage{algorithm}
\usepackage{stackengine}
\usepackage{amssymb}
\usepackage{caption}
\usepackage{amsmath}
\usepackage{booktabs}
\usepackage{bold-extra}
\usepackage{float}
\usepackage{bm}
\usepackage{empheq} 
\usepackage{makecell} 
\usepackage{changepage}
\usepackage{tabularx, booktabs}
\DeclareCaptionLabelFormat{andtable}{#1~#2  \&  \tablename~\thetable}

\algrenewcommand{\alglinenumber}[1]{\tiny#1:}

\newcommand*\widefbox[1]{\fbox{\hspace{1em}#1\hspace{1em}}} 

\begin{document}
\title{Deducing Closed-Form Expressions for Bright-Solitons in Strongly Magnetized Plasmas with Physics Informed Symbolic Regression (PISR)}
%
%
\author{Edward Finkelstein\inst{1}
\thanks{Supported by the \href{https://www.smartscholarship.org/smart}{SMART} program}}
\authorrunning{Edward F.}
%
\institute{Naval Research Lab\\
4555 Overlook Ave., SW, Washington, DC\\ \email{edward.finkelstein3.ctr@us.navy.mil}}

\maketitle              
\begin{abstract}
This paper presents a novel approach to finding analytical approximations for bright-soliton solutions in strongly magnetized plasmas. We leverage Physics-Informed Symbolic Regression (PISR) to discover closed-form expressions for the vector potential and number density profiles, governed by a reduced-order model derived from Maxwell-fluid equations.  The PISR framework combines symbolic regression with physics-based constraints, boundary conditions, and available simulation data to guide the search for solutions. We demonstrate the effectiveness of the approach by rediscovering approximate solutions consistent with previously published numerical results, showcasing the potential of PISR for reducing simulation costs of reduced-order models in plasma physics.
\end{abstract}

\section{Introduction}

Bright solitons are localized, self-reinforcing wave packets that maintain their shape and speed during propagation due to a balance between nonlinear effects and dispersion \cite{Kivshar1998}. They are of significant interest in various fields, including nonlinear optics \cite{Agrawal2007}, Bose-Einstein condensates \cite{Dalfovo1999}, and plasma physics \cite{Pegoraro2000Nonlinear}. In plasma physics, understanding and predicting the behavior of bright solitons is crucial for applications such as laser-plasma acceleration \cite{Esarey2009} and inertial confinement fusion \cite{Atzeni2004}.

Traditional methods for studying solitons often rely on numerical simulations \cite{Taflove1995ComputationalET}, \cite{Vlasov1961}, which can be computationally expensive, especially for long-time dynamics or high-dimensional systems \cite{Dawson1983}. While numerical simulations provide valuable insights, they can sometimes offer limited physical understanding due to the complexity of the underlying algorithms and the vast amount of data generated \cite{Winsberg2010}. Analytical solutions, while desirable for their interpretability and predictive power \cite{Whitham1999}, are often difficult to obtain, particularly for complex plasma models with multiple interacting species and nonlinear effects \cite{Davidson1972}. This motivates the development of alternative approaches that can provide accurate and interpretable approximations while maintaining computational efficiency and physical insight.

This paper introduces a Physics-Informed \cite{Karniadakis2021} Symbolic Regression (PISR) approach for discovering closed-form expressions for bright-soliton solutions in strongly magnetized plasmas.  PISR combines the power of symbolic regression with physics-based constraints, boundary conditions, and available simulation data to guide the search for solutions.  This approach allows us to obtain analytical approximations that are consistent with the underlying physics and can be easily interpreted and manipulated.

\subsection{Symbolic Regression}
In this work, we employ symbolic regression (SR) to solve a system of equations, denoted by $\vec{F}\left(\vec{f}\left(\vec{x}\right), \vec{x}\right) = 0$.  The goal of SR is to identify the unknown functions $\vec{f}\left(\vec{x}\right) = \left\{f_1\left(\vec{x}_1\right), f_2\left(\vec{x}_2\right), \ldots, f_N\left(\vec{x}_N\right)\right\}$.  Here, $\vec{x}$ represents the independent variables in the system $\vec{F}$, and each $\vec{x}_i$ is a subset (possibly improper) of $\vec{x}$, indicating the specific independent variables on which the function $f_i$ depends.

\par The nodes within each symbolic expression in $\vec{f}$ fall into one of the following three categories \cite{lacava2021contemporarysymbolicregressionmethods}:
\begin{itemize}
\item \textbf{Unary operators: } Operators with arity 1, such as $\cos$, $\sin$, $\exp$, $\ln$, $\tanh$, etc.
\item \textbf{Binary operators: } Operators with arity 2, such as $+$, $-$, $*$, $\div$, etc.
\item \textbf{Leaf Nodes: } The individual features $\vec{x} = \{x_1, x_2, \ldots,x_{N}\}$ and constant tokens.  These constants can be further refined using non-linear optimization techniques like L-BFGS \cite{doi:10.1137/0916069} or Levenberg-Marquardt \cite{83b09f23-b20e-3617-8f72-24765b713f7b} \cite{doi:10.1137/0111030}.
\end{itemize}
A key advantage of symbolic regression lies in the inherent interpretability of the resulting functional forms, a characteristic often lacking in other machine learning methods. This makes SR particularly valuable in domains where understanding the underlying relationships is crucial, such as health, law, and the natural sciences \cite{lacava2021contemporarysymbolicregressionmethods}.

\subsection{Symbolic Differentiation}
Symbolic differentiation is a computer algebra technique for analytically computing derivatives \cite{10.1007/978-3-642-57201-2_12} using rules such as the chain, product, and quotient rules. This approach falls under the broader category of ``computer algebra'' \cite{tan2000symbolicc}, readily available in software packages like Mathematica \cite{Mathematica}, SymPy \cite{10.7717/peerj-cs.103}, Maxima \cite{maxima}, and Maple \cite{maple}.  However, as noted in \cite{PredragV2001}, many symbolic differentiation implementations rely on tree or graph data structures to represent formulae and their derivatives. While versatile, these data structures may be suboptimal for symbolic regression, which demands minimal memory allocation and maximal computational speed for efficient exploration of the search space \cite{virgolin2022symbolicregressionnphard}.

\par Therefore, this work implements the array-based symbolic differentiation method developed in \cite{PredragV2001}, along with its in-situ simplification routines.  This approach minimizes memory allocation by avoiding the creation of new expression trees. Furthermore, since expressions are represented in prefix/postfix notation (following the grammars and algorithms developed in \cite{finkelstein2024generalizedfixeddepthprefixpostfix}), the method eliminates the need for infix-to-prefix/postfix conversion, contributing to improved efficiency. This efficiency is crucial when scaling to higher-dimensional input datasets \cite{Kamienny2023DeepGS}.

\subsection{Symbolic Simplification}

In this work, symbolic simplification refers to the algorithmic reduction of node count within candidate solution expressions, $\vec{f}$, to minimize computational latency. The goal is to reduce the floating-point operations required to evaluate $\vec{F}\left(\vec{f}\right)$, regardless of whether these equations represent ordinary differential equations, partial differential equations, or algebraic constraints.

\par Symbolic simplification has a rich history within computer algebra systems (CAS) and expert systems \cite{Buchberger1985}, \cite{Davenport1988}. Early CAS, like Maxima \cite{maxima}, \cite{Moses1971} and Reduce \cite{Hearn1973}, invested heavily in simplification algorithms using rule-based approaches, pattern matching, and canonicalization techniques. However, symbolic simplification is fundamentally challenging. Determining the ``simplest'' form of an expression is generally undecidable, and many simplification problems are NP-hard \cite{Plaisted1978}. Furthermore, the notion of ``simplicity'' is subjective and application-dependent \cite{Stoutemyer1985}.

\par Expert systems have also employed symbolic simplification to manage expression complexity during problem-solving \cite{Feigenbaum1963}. However, maintaining a comprehensive and consistent set of simplification rules can be challenging, and system performance can be sensitive to rule application order.

\par For our purposes, we adopt a pragmatic approach. Our objective is not to achieve absolute simplicity, but to minimize expression size and computational cost during evaluation. This is particularly critical in symbolic regression, where candidate solutions are repeatedly evaluated over large datasets. By reducing the operations needed to evaluate each expression, we significantly improve search efficiency \cite{PredragV2001}. We leverage efficient in-situ simplification routines, such as those in \cite{PredragV2001}, to achieve this with minimal overhead.

\subsection{Expression Representations: Polish and Reverse Polish Notation}

Symbolic expressions can be represented in infix, prefix (Polish notation), and postfix (Reverse Polish notation). While infix notation is the most familiar (e.g., \texttt{2 + 2}), prefix and postfix notations offer advantages for symbolic manipulation and evaluation.

\begin{itemize}
\item \textbf{Polish Notation (Prefix):} Operators precede their operands (e.g., \texttt{+ 2 2}).
\item \textbf{Reverse Polish Notation (Postfix):} Operators follow their operands (e.g., \texttt{2 2 +}).
\end{itemize}

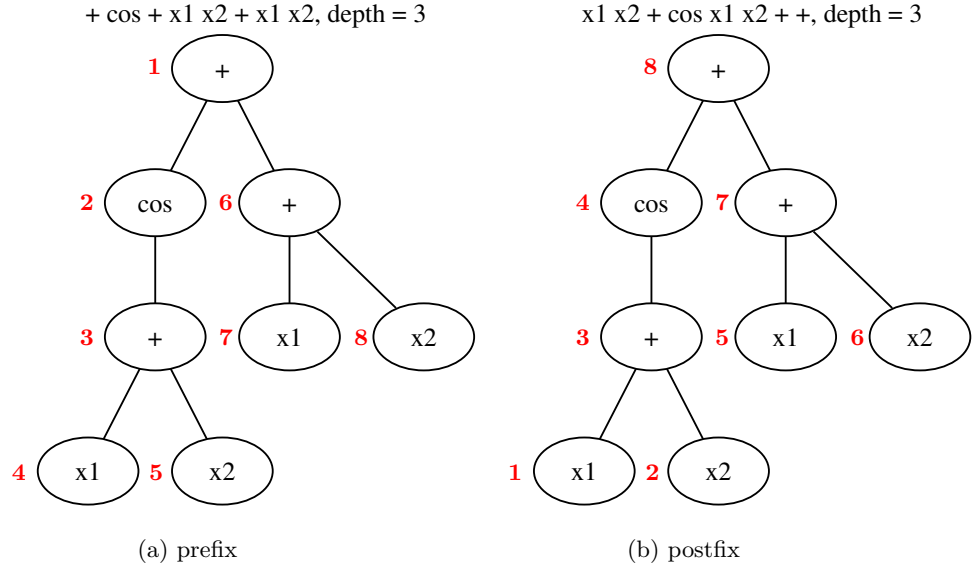
\begin{figure}[ht]
\centering
\begin{subfigure}[b]{0.4\textwidth}
\centering
\begin{tikzpicture}
\node[text width = 6 cm, align = center] at (0,0) {\includegraphics[width=\linewidth, keepaspectratio]
{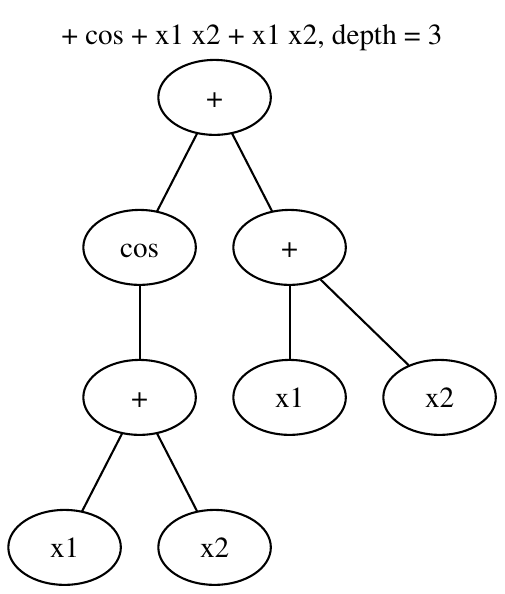}};
\node at (-1.35, 2.4) {\textcolor{red}{\textbf{1}}};
\node at (-2.25, 0.6) {\textcolor{red}{\textbf{2}}};
\node at (-2.25, -1.2) {\textcolor{red}{\textbf{3}}};
\node at (-3.15, -3) {\textcolor{red}{\textbf{4}}};
\node at (-1.325, -3) {\textcolor{red}{\textbf{5}}};
\node at (-0.4, 0.6) {\textcolor{red}{\textbf{6}}};
\node at (-0.4, -1.2) {\textcolor{red}{\textbf{7}}};
\node at (1.4, -1.2) {\textcolor{red}{\textbf{8}}};
\end{tikzpicture}
\caption{prefix}
\label{subfig:prefix_tree_example}
\end{subfigure}
\hspace{1.5cm}
\begin{subfigure}[b]{0.4\textwidth}
\centering
\begin{tikzpicture}
\node[text width = 6 cm, align = center] at (0,0) {\includegraphics[width=\linewidth, keepaspectratio]{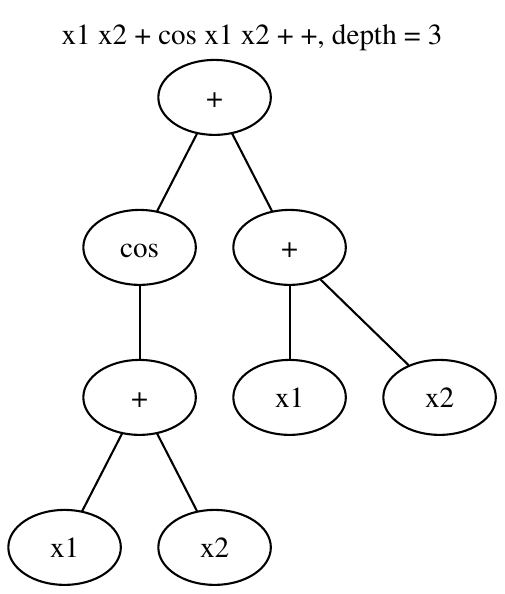}};
\node at (-1.35, 2.4) {\textcolor{red}{\textbf{8}}};
\node at (-2.25, 0.6) {\textcolor{red}{\textbf{4}}};
\node at (-2.25, -1.2) {\textcolor{red}{\textbf{3}}};
\node at (-3.15, -3) {\textcolor{red}{\textbf{1}}};
\node at (-1.325, -3) {\textcolor{red}{\textbf{2}}};
\node at (-0.4, 0.6) {\textcolor{red}{\textbf{7}}};
\node at (-0.4, -1.2) {\textcolor{red}{\textbf{5}}};
\node at (1.4, -1.2) {\textcolor{red}{\textbf{6}}};
\end{tikzpicture}
\caption{postfix} \label{subfig:postfix_tree_example}
\end{subfigure}
\caption{Expression Tree Representations}
\label{fig:expression_trees}
\end{figure}

\par This work leverages Reverse Polish notation for representing symbolic expressions. We have empirically found this representation to be more efficient, particularly when combined with the fixed-depth grammar framework developed in \cite{finkelstein2024generalizedfixeddepthprefixpostfix}. This choice facilitates the efficient symbolic differentiation and simplification essential for our Physics-Informed Symbolic Regression (PISR) approach.

\section{Related Work}

The field of discovering and solving differential equations via symbolic methods has gained considerable momentum. A key distinction exists between the \emph{forward problem}, where the governing equation $\vec{F}$ is known and the solution $\vec{f}$ is sought, and the \emph{inverse problem}, where data from the solution $\vec{f}$ is available, and the objective is to discover the governing equation $\vec{F}$. This work focuses on addressing the \emph{forward} problem.

\par In \cite{oh2023geneticprogrammingbasedsymbolic}, the authors tackle the forward problem, employing the Bingo framework \cite{10.1145/3520304.3534031} with PyTorch-based automatic differentiation \cite{paszke2019pytorchimperativestylehighperformance} to solve Euler-Bernoulli and Poisson equations subject to boundary constraints. They do not consider initial condition constraints and observe that PyTorch differentiation exhibits suboptimal performance compared to direct function evaluation.

\par The authors of \cite{Manti_2024} address the inverse problem, distilling solutions for Poisson, Euler Elastica, and linear elasticity boundary-value problems. They utilize a custom library of discrete-exterior calculus operators implemented with DEAP \cite{10.5555/2503308.2503311}, Ray \cite{10.5555/3291168.3291210}, JAX \cite{jax2018github}, and pygmo \cite{Biscani2020}. This framework leverages strongly typed genetic programming, parallelization, automatic differentiation, and LBFGS optimization. While powerful, its application to dynamical system distillation is stated to be forthcoming \cite{desilva2020}.

\par The Sparse Identification of Nonlinear Dynamical Systems (SINDy) framework \cite{Kaptanoglu2022}, \cite{desilva2020} aims to efficiently solve the inverse problem. The authors of \cite{Kaptanoglu2022} assume a governing equation of the form $\sum_{k=1}^{p} \theta_{k}(\vec{x}) \xi_{k}$ for the dynamical system $d\vec{x}(t)/dt = \vec{f}(\vec{x}(t))$, where $\theta_k$ are pre-specified basis functions and $\xi_{k}$ are the corresponding coefficients. Using PySINDy and leveraging this sparsity assumption, they can efficiently discover $\vec{f}$. However, this approach relies heavily on the choice of basis functions. Consequently, it is best suited for systems where the form of the governing dynamical system equation $\vec{F}$ is partially known and appears primarily applicable to strictly \emph{dynamical} systems.

\par The authors of \cite{sun2023symbolicphysicslearnerdiscovering} introduce the Symbolic Physics Learner (SPL) framework, evaluating it on inverse problems involving experiments of chaotic systems like the double pendulum and the 3D Lorenz system. Their SPL framework endeavors to discover governing equations from observed data. The acknowledged limitations of \cite{sun2023symbolicphysicslearnerdiscovering}, specifically the lack of multi-threading and inaccurate state-derivative approximation, are addressed in this work through a multi-threaded C++ Eigen \cite{eigenweb} symbolic regression framework.  This framework leverages the symbolic differentiation method of \cite{PredragV2001} for exact derivative computation and enhanced efficiency.

\par Our PISR application most closely aligns with the work of Das et al. \cite{Das2025}. In \cite{Das2025}, the authors employ a physics-informed neural network to approximate solutions of differential equations arising in the mathematical modeling of arterial blood flow. They subsequently use PySR to obtain symbolic equations that best fit these neural network solutions.  We suggest that this two-stage methodology may introduce inefficiencies. Our method, which directly applies symbolic regression to the differential equation in C++, has the potential to be more efficient and accurate by directly addressing the forward problem.  However, it is important to note that the authors of \cite{Das2025} achieved non-trivial solutions without relying on simulation data within their loss function\footnote{This detail was confirmed in a private correspondence.}.

\section{Problem Formulation: Bright Solitons in Magnetized Plasmas}

We consider a warm, magnetized, quasi-neutral plasma described by the Maxwell-fluid equations.  As shown in \cite{FengLaserPaper}, under the right-hand circular polarization and quasi-neutrality approximations, the system can be reduced to a set of coupled ordinary differential equations governing the vector potential and number density profiles of bright solitons.

The starting point is the Maxwell-Fluid Equations:

\begin{align}
    \Delta \varphi  &= 4\pi e(n_e- n_i), \label{eq:maxwell1} \\[0.3cm]
    \Delta \boldsymbol A- \dfrac{1}{c^2}\dfrac{\partial^2\boldsymbol A}{\partial t^2}- \dfrac{1}{c}\dfrac{\partial }{\partial t}\nabla \varphi  &= \dfrac{4\pi e}{c}(n_e\boldsymbol v_e- n_i\boldsymbol v_i)\ldotp, \label{eq:maxwell2} \\[0.3cm]
    \dfrac{\partial }{\partial t}(\boldsymbol p_s + \dfrac{q_s}{c}\boldsymbol A) &= - \nabla (q_s\varphi  + \gamma_sm_sc^2) \nonumber \\  &\qquad + \boldsymbol v_s \times [\nabla  \times (\boldsymbol p_s + \dfrac{q_s}{c}\boldsymbol A)] \nonumber \\  &\qquad + \dfrac{q_s}{c}\boldsymbol v_s \times \boldsymbol B_0- m_s\boldsymbol v_{ts}^2\nabla \text{ln}n_s \label{eq:maxwell3}
\end{align}

To simplify this system, the following assumptions are made (see \cite{FengLaserPaper} for details):

\begin{itemize}
\item   \textbf{Right-Hand Circular Polarization:}
    \begin{align}
        \boldsymbol{A} = \left\{ A_x(y), 0, A_z(y) \right\}, \;\; & \;\;\, \boldsymbol{p} = \left\{ p_{s,x}(y), 0, p_{s,z}(y) \right\} \nonumber \\[0.25cm]
        A_z + iA_x &\sim a(y)\,\text{exp}(i\omega t) \nonumber \\[0.3cm]
        p_{s,z} + i p_{s,x} &\sim p_{s\bot}(y)\,\text{exp}(i\omega t) \nonumber \\[0.3cm]
        \varphi  &\sim \varphi (y) \nonumber
    \end{align}
\item   \textbf{Quasi-Neutral Approximation:} $n_e = n_i = n$
\end{itemize}

These assumptions lead to the following Reduced Order Model (ROM):

\begin{empheq}[box=\widefbox]{align}
    &\dfrac{\partial^2}{\partial \xi^2}\left((\sinh(u(\xi)) - \alpha \tanh(u(\xi))) + \omega^2(\sinh(u(\xi)) - \alpha \tanh(u(\xi))\right) \nonumber \\ &- n(\xi)\left(\dfrac{\tanh(u(\xi)) + \rho_i\sinh(u(\xi))}{1 + \rho_i\alpha }\right) = 0 \label{eq:rom1} \\[0.3cm]
    &\left(\rho_iv_{te}^2 + v_{ti}^2\right)\ln (n(\xi)) - \rho_i(1 - \cosh(u(\xi))) + \dfrac{1}{2}\rho_i\,\alpha\, \tanh^2(u(\xi)) \nonumber \\ &-\dfrac{\rho_i^2(\sinh(u(\xi)) - \alpha \tanh(u(\xi)))^2}{ 2\cdot(1 + \rho_i\alpha )} \approx 0 \label{eq:rom2}
\end{empheq}

where:

\begin{align}
    &\rho_i = \dfrac{m_e}{m_i} \approx \dfrac{1}{1836}, \quad (\text{mass-ratio})\nonumber\\[0.2cm]
    &\alpha = -\dfrac{q_e B_0}{m_e c}\cdot\dfrac{1}{\omega} = \dfrac{\omega_{ce}}{\omega}, \quad (\text{frequency ratio}) \nonumber\\[0.2cm]
     &a(\xi) =\sinh\left(u(\xi)\right)-\alpha\cdot\gamma_{0}\cdot\tanh\left(u(\xi)\right), \quad (\propto \, A)  \nonumber\\[0.2cm]
     &\gamma_0 = \dfrac{1}{\sqrt{1-\dfrac{V^2}{c^2}}}, \quad \xi = x - Vt, \nonumber\\[0.2cm]
     &\text{$V$: velocity of bright-soliton} \nonumber\\
     &\text{$c$: speed of light} \nonumber
\end{align}

The goal of this work is to find expressions for $u(\xi)$ and $n(\xi)$ that satisfy the ROM equations \eqref{eq:rom1} and \eqref{eq:rom2}.

\section{Methods}

Our approach uses Physics-Informed Symbolic Regression (PISR) to discover closed-form expressions for $u(\xi)$ and $n(\xi)$. The key components of our PISR framework are:

\begin{enumerate}
\item  \textbf{Symbolic Regression:} We use a symbolic regression algorithm to generate candidate expressions for $u(\xi)$ and $n(\xi)$. The algorithm searches through a space of possible expressions constructed from a predefined set of operators (e.g., +, -, \*, /, sin, cos, exp, etc.) and variables (e.g., $\xi$). The fixed-depth grammar developed in \cite{finkelstein2024generalizedfixeddepthprefixpostfix} is used to build expressions.

\item  \textbf{Physics-Based Loss Function:} We define a loss function that penalizes expressions that do not satisfy the governing equations and boundary conditions. The loss function is composed of several terms:
\begin{itemize}
\item   \textbf{Equation Loss:} Measures the residual of the ROM equations \eqref{eq:rom1} and \eqref{eq:rom2} when the candidate expressions for $u(\xi)$ and $n(\xi)$ are substituted into them.
\item   \textbf{Boundary Condition Loss:} Enforces the boundary conditions on the solution.
\item   \textbf{Data Loss:} Measures the difference between the candidate solutions and available simulation data.
\item   \textbf{Symmetry Loss:} Enforces symmetry conditions on the solution (e.g., $n(x) = n(-x)$).
\end{itemize}

\item  \textbf{Symbolic Differentiation:} We use symbolic differentiation to compute the derivatives required in the ROM equations and the loss function. This ensures accurate and efficient evaluation of the loss function. The method from \cite{PredragV2001} is used for symbolic differentiation.

\item  \textbf{Optimization:} We use an optimization algorithm to minimize the loss function and find the expressions for $u(\xi)$ and $n(\xi)$ that best satisfy the physics-based constraints.
\end{enumerate}

The overall optimization problem can be formulated as:

\begin{equation}
\min_{u(\xi), n(\xi)} \mathcal{L} = \mathcal{L}_{\text{equation}} + \mathcal{L}_{\text{boundary}} + \mathcal{L}_{\text{data}} + \mathcal{L}_{\text{symmetry}}
\end{equation}

where $\mathcal{L}$ is the total loss function, and the subscripts indicate the individual loss terms.

\section{Experimental Setup}

To validate the PISR framework, we applied it to the bright-soliton problem described in Section 3. We used the following setup\footnote{replacing the symbol $\xi$ with $x$ for notational convenience in the following.}

\begin{itemize}
\item   \textbf{Data:} We used the numerical simulation data from Figure 10 of \cite{FengLaserPaper} for $u(x)$ and $n(x)$. The data consists of 127 points.
\item   \textbf{Parameters:} We used the same parameter values as in \cite{FengLaserPaper}: $\rho_i = 1/1836$, $\alpha = 0.4$, $v_{te} = 0.05c$, $v_{ti} = 0.001c$, $n_0=1$, and $\omega^2 = 0.64 \cdot n(x)$.
\item   \textbf{Loss Function:} The loss function includes the equation loss, boundary condition loss, data loss, and symmetry loss, as described in Section 4.
\item   \textbf{Symbolic Regression:} We used brute-force and simulated annealing algorithms to generate candidate expressions for $u(x)$ and $n(x)$.
\item   \textbf{Operators:}
	\begin{itemize}
		\item Unary Operators: \textbf{-}, $\mathbf{log}$, $\mathbf{exp}$, $\mathbf{cos}$, $\mathbf{sin}$, $\mathbf{sqrt}$, $\mathbf{asin}$, $\mathbf{acos}$, $\mathbf{tanh}$, $\mathbf{sech}$
		\item Binary Operators: \textbf{+}, \textbf{-}, \textbf{*}, \textbf{/}, $\bm{\wedge}$
	\end{itemize}
\item   \textbf{Independent Variable:} The independent variable here is $x$.
\item	\textbf{Trivial Solution Threshold:} $\text{Var}(u(x)) = \text{Var}(n(x)) = \text{Var}(\mathrm{d}u(x)/dx) = \text{Var}(\mathrm{d}n(x)/dx) \geq \bm{10^{-3}}$
\item   \textbf{Implementation:} The framework is implemented in C++ with Boost \cite{BoostLibrary} and Eigen \cite{eigenweb} dependencies, as well as LBFGS++ \cite{lbfgspp} as an alternative constant-fitting option to \href{https://libeigen.gitlab.io/eigen/docs-nightly/unsupported/classEigen_1_1LevenbergMarquardt.html}{Eigen's Levenberg-Marquardt implementation}. The computer we run on is shown in section \ref{sec:computer_info}.
\end{itemize}
The specific equations we used in the loss function are as follows:
\begin{itemize}
\item   \textbf{Feng's Original Two Equations:}
    \begin{align}
        &\dfrac{\partial^2}{\partial x^2}\left((\sinh(u(x)) - \alpha \tanh(u(x))) + \omega^2(\sinh(u(x)) - \alpha \tanh(u(x))\right) \nonumber \\ &- n(x)\left(\dfrac{\tanh(u(x)) + \rho_i\sinh(u(x))}{1 + \rho_i\alpha }\right) = 0 \label{eq:loss1} \\[0.3cm]
        &\left(\rho_iv_{te}^2 + v_{ti}^2\right)\ln (n(x)) - \rho_i(1 - \cosh(u(x))) + \dfrac{1}{2}\rho_i\,\alpha\, \tanh^2(u(x)) \nonumber \\ &-\dfrac{\rho_i^2(\sinh(u(x)) - \alpha \tanh(u(x)))^2}{ 2\cdot(1 + \rho_i\alpha )} \approx 0 \label{eq:loss2}
    \end{align}
\item   \textbf{Boundary + Symmetry Conditions:}
    \begin{align}
        &a(x_{\text{min}}) =  \sinh\left(u(x_{\text{min}})\right)-\alpha\cdot\gamma_{0}\cdot\tanh\left(u(x_{\text{min}})\right) = 0 \label{eq:loss3} \\[0.3cm]
        &a(x_{\text{max}}) = \sinh\left(u(x_{\text{max}})\right)-\alpha\cdot\gamma_{0}\cdot\tanh\left(u(x_{\text{max}})\right) = 0 \label{eq:loss4} \\[0.3cm]
        &\dfrac{\partial a(x_{\text{min}})}{\partial x} =\left|\left(\cosh\left(u\right) - {\alpha}{\gamma}_{0} \operatorname{sech}^{2}\left(u\right)\right)\dfrac{\partial u}{\partial x}\right|_{x = x_{\text{min}}} = 0 \label{eq:loss5} \\[0.3cm]
        &\dfrac{\partial a(x_{\text{max}})}{\partial x} =\left|\left(\cosh\left(u\right) - {\alpha}{\gamma}_{0} \operatorname{sech}^{2}\left(u\right)\right)\dfrac{\partial u}{\partial x}\right|_{x = x_{\text{max}}} = 0 \label{eq:loss6} \\[0.3cm]
        &\left|n(x) - n(-x)\right| = 0 \label{eq:loss7}
    \end{align}
\item   \textbf{Feng's Simulation Data\footnote{We multiply the second equation \ref{eq:loss9} by 10 because Feng's simulated $a(x)$ is approximately an order of magnitude smaller than $\text{Density}(x)$; see Figure 10 of \cite{FengLaserPaper}.}:}
    \begin{align}
        &\left|\text{Density}(x) - \mathcal{D}_1\right| = \left|\left(\dfrac{n(x)}{n_0} - 1\right) - \mathcal{D}_{1}\right| = 0 \label{eq:loss8} \\[0.3cm]
        &10\cdot  \left|a(x) - \mathcal{D}_2\right| = 10\cdot \left|\left(\sinh\left(u(x)\right)-\alpha\cdot\gamma_{0}\cdot\tanh\left(u(x)\right)\right) - \mathcal{D}_{2}\right| = 0 \label{eq:loss9}
    \end{align}
\end{itemize}
\section{Results}
Using our PISR approach, we discovered the following expressions for $u(x)$ and $n(x)$:
\begin{empheq}[box=\widefbox]{align*}
    u(x) &\approx \dfrac{\tanh{\left(\tanh{\left(\mathrm{sech}{\left(x \right)} \right)} \right)}}{- \dfrac{\tanh{\left(x \right)}}{e} - 6.434} \\[0.225cm]
    n(x) &=\mathrm{sech}{\left(3.235\cdot \mathrm{sech}^{\tanh{\left(2 \right)}}{\left(x \right)} \right)}\\[0.15cm]
\gamma_0 &\approx 5.22145 \quad \text{(fitted automatically)}
\end{empheq}

These expressions provide an approximate analytical solution to the bright-soliton problem. Figure 2 shows a comparison between the discovered expressions and the numerical simulation data from \cite{FengLaserPaper}.

\begin{figure}[ht]
    \centering
    \includegraphics[scale=0.6]{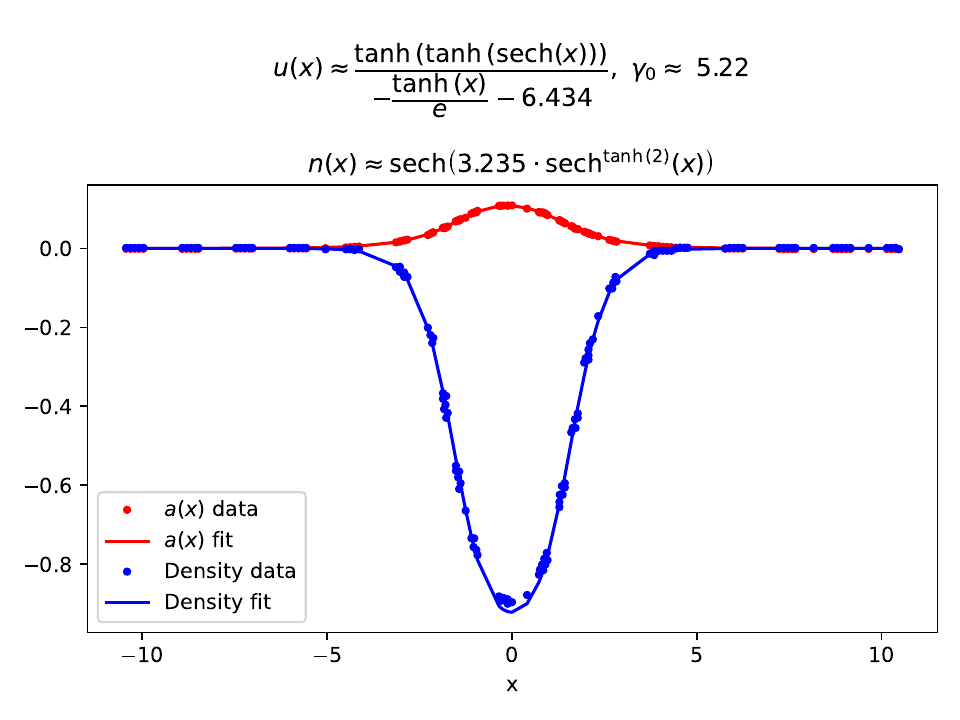}
    \caption{Comparison of Discovered Expressions and Numerical Simulation Data.  $a(x) =\sinh\left(u(x)\right)-\alpha\cdot\gamma_{0}\cdot\tanh\left(u(x)\right)$}
    \label{fig:comparison}
\end{figure}

Table \ref{tab:sne} shows the squared-norm error (SNE) for each term in the loss function. The total SNE is 0.0602, indicating a good agreement between the discovered expressions and the physics-based constraints and data.

\begin{table}[H]
    \centering
    \begin{tabular}{|c|c|c|} \hline
        \textbf{Equation \#} & \textbf{Description} & \makecell{\textbf{Squared Norm Error} \\ \textbf{(SNE)}} \\[0.2cm] \hline
        \ref{eq:loss1} & Feng Equation 1 & 0.0209 \\ \hline
        \ref{eq:loss2} & Feng Equation 2 & 0.0 \\ \hline
        \ref{eq:loss3} & Boundary Condition 1 for $a(x)$ & 0.0 \\ \hline
        \ref{eq:loss4} & Boundary Condition 2 for $a(x)$ & 0.0 \\ \hline
        \ref{eq:loss5} & Boundary Condition 3 for $a(x)$ & 0.0 \\ \hline
        \ref{eq:loss6} & Boundary Condition 4 for $a(x)$ & 0.0 \\ \hline
        \ref{eq:loss7} & Symmetry Condition 1 for $n(x)$ & 0.0 \\ \hline
        \ref{eq:loss8} & Data for density & 0.0214 \\ \hline
        \ref{eq:loss9} & Data for $a(x)$ & 0.0179 \\ \hline
    \end{tabular}
    \caption{Squared Norm Error for Each Term in the Loss Function. The number of data-points is 127; thus, the MSE is computed by dividing each term in this table by 127.}
    \label{tab:sne}
\end{table}

\section{Discussion and Potential for Improvement}

The results demonstrate the effectiveness of our PISR approach for discovering analytical approximations for bright-soliton solutions in strongly magnetized plasmas. The discovered expressions capture the essential physics and exhibit good agreement with numerical simulation data.

\par Despite these successes, several areas offer opportunities for improvement:

\begin{enumerate}
    \item \textbf{Optimization of System Construction:}\label{item:optim_sys} The current approach constructs the system of equations, $\vec{F}$, by symbolically generating $\vec{f}$, simplifying $\vec{F}$ symbolically, and then evaluating $\vec{F}$ numerically.  Efficiency could be improved by identifying and reusing common sub-expressions within $\vec{F}$, a technique analogous to common subexpression elimination in compiler optimization \cite{10.5555/286076}.

    \item \textbf{Dynamic Simplification:}\label{item:dyn_simp} The simulated annealing approach currently maintains a fixed expression depth throughout the optimization process and doesn't allow simplification during perturbation.  Enabling symbolic simplification of $\vec{f}$ during the search, using a temporary vector to store the evaluation-ready form of $\vec{f}$ separately from the form being perturbed, would likely improve performance.

    \item \textbf{Robust Trivial Solution Rejection:}\label{item:rej_triv_sol}  The current criteria for rejecting trivial solutions include checks for variable presence, variance, and a threshold on the maximum of each partial derivative.  A more robust approach would incorporate the \emph{median} of the partial derivatives in addition to the maximum, as the median is less sensitive to outliers \cite{559660}.

    \item \textbf{Equation Scaling and Weighting:}\label{item:eqn_scale_weight}  The equations within the system $\vec{F}$ (Eqs. \ref{eq:loss1} - \ref{eq:loss9}) may have disparate magnitudes, potentially leading to unequal weighting during optimization.  Future work should explore scaling each equation to ensure equal contribution to the loss function. Furthermore, applying a ``squishification" function, such as $\tanh$, or more advanced rescaling techniques to the right-hand side of $\vec{F}$ could mitigate the impact of remaining order-of-magnitude differences between terms within individual equations.

	\item \textbf{Data Sensitivity Analysis:}\label{item:data_sens} A key area for future investigation is the sensitivity of the PISR-discovered solutions to the quantity of training data. Performing a downsampling analysis, where the number of data points used to train the PISR model is systematically reduced, would provide valuable insights into the robustness and generalizability of the approach. A potential metric to track would be the total SNE achieved after a fixed training time (e.g., one hour) as a function of the number of data points used for training. Additionally, tracking changes in the discovered symbolic expressions themselves would indicate the stability of the solutions.

	\item \textbf{Analytical Refinement via Perturbation:}\label{item:perturb_refine} The PISR-discovered expressions could serve as a valuable starting point for obtaining more accurate or even exact analytical solutions. Specifically, the PISR results could be used as a zeroth-order term in a perturbative expansion of the governing equations. Namely, substituting the PISR expressions into the governing system of equations and analyzing the resulting residual (remainder term) may enable the derivation of a reduced expression for the higher-order terms in the expansion and a refined analytical solution.
\end{enumerate}

\par Nevertheless, we believe this work successfully demonstrates the potential of PISR for solving systems of differential and algebraic equations in plasma physics. Building on this foundation, and recognizing the limitations of current surrogate modeling techniques (highlighted in \cite{ohana2024well}), future research should focus on developing more scalable and efficient PISR algorithms for tackling complex, high-resolution problems.

\bibliographystyle{splncs04}
\bibliography{PrefixPostfixPaper}
\newpage
\appendix
\section{Computer Info for Reproducibility}\label{sec:computer_info}
All tests in this paper were run on a Windows Computer. The following command was executed on the computer in the Windows PowerShell application:
\begin{verbatim}
Get-ComputerInfo | Select-Object -Property @{N='OSName';E={$_.OsName}},
@{N='OSVersion';E={$_.OsVersion}},
@{N='OsArchitecture';E={$_.OsArchitecture}},
@{N='CsNumberOfLogicalProcessors';E={$_.CsNumberOfLogicalProcessors}},
@{N='CsNumberOfProcessors';E={$_.CsNumberOfProcessors}},
@{N='CsProcessors';E={$_.CsProcessors}},
@{N='CsTotalPhysicalMemory';E={$_.CsTotalPhysicalMemory}},
@{N='BiosName';E={$_.BiosName}},
@{N='BiosVersion';E={$_.BiosVersion}},
@{N='CsModel';E={$_.CsModel}}
\end{verbatim}

This command yielded the following output:
\begin{verbatim}
OSName                      : Microsoft Windows 11 Enterprise
OSVersion                   : 10.0.26100
OsArchitecture              : 64-bit
CsNumberOfLogicalProcessors : 8
CsNumberOfProcessors        : 1
CsProcessors                : Microsoft.PowerShell.Commands.Processor
CsTotalPhysicalMemory       : 34023538688
BiosName                    : 1.1.2
BiosVersion                 : DELL   - 1072009
CsModel                     : Precision 5820 Tower
\end{verbatim}
\end{document}